\documentclass[sigconf]{acmart}
\usepackage{enumitem}
\usepackage{tikz}
\usetikzlibrary{fit,positioning,arrows.meta,calc,automata,shapes.geometric}
\usepackage{subcaption}

\AtBeginDocument{%
  }

\setcopyright{acmlicensed}
\copyrightyear{2026}
\acmYear{2026}
\setcopyright{cc}
\setcctype{by}
\acmConference[MODELS 2026]{ACM/IEEE 29th International Conference on Model Driven Engineering Languages and Systems}{October 04--09, 2026}{Málaga, Spain}
\acmBooktitle{ACM/IEEE 29th International Conference on Model Driven Engineering Languages and Systems (MODELS 2026), October 04--09, 2026, Málaga, Spain}
\acmDOI{10.1145/3822455.3838777}
\acmISBN{979-8-4007-2809-9/2026/10}

\begin{document}
\title{ATLAS: Discovering Agent Strategies through LLM-Guided Abstraction and Automata Learning}

%%
% \author{Anonymous Author(s)}
%     \affiliation{
%     \institution{Institution}
%     \city{City}
%     \country{Country}
% }
\author{Ignacio D. Lopez-Miguel}
\affiliation{%
  \institution{TU Wien}
  \city{Vienna}
  \country{Austria}
}

\author{Andreas Happe}
\affiliation{%
  \institution{TU Wien}
  \city{Vienna}
  \country{Austria}
}

\author{J{\"u}rgen Cito}
\affiliation{%
  \institution{TU Wien}
  \city{Vienna}
  \country{Austria}
}

\author{Ezio Bartocci}
\affiliation{%
  \institution{TU Wien}
  \city{Vienna}
  \country{Austria}
}

\author{Bettina Könighofer}
\affiliation{%
  \institution{TU Graz}
  \city{Graz}
  \country{Austria}
}

\author{Martin Tappler}
\affiliation{%
  \institution{TU Wien}
  \city{Vienna}
  \country{Austria}
}

% \renewcommand{\shortauthors}{Lopez-Miguel et al.}

\begin{abstract}
Large Language Model (LLM)-based agents are increasingly used for complex tasks such as software testing and cybersecurity assessment. While these agents demonstrate impressive capabilities, their behavior is difficult to understand, explain, and analyze. Existing evaluations focus mainly on task success and execution traces, offering limited insight into the strategies employed by the agent. We present ATLAS (Automata Learning for Agent Trajectory Analysis and Strategy Discovery), an approach for recovering interpretable behavioral models from agent trajectories. ATLAS combines trace abstraction with automata learning to infer finite-state models that capture observed agent-environment interaction strategies. These models provide human-interpretable insights and support automated analyses of recurring behaviors, decision points, successful task-completion paths, and failure loops. As a proof of concept, we apply ATLAS to trajectories generated by an LLM-based penetration-testing agent. The resulting models expose high-level behavioral strategies for exploiting vulnerable machines that are difficult to identify from raw execution traces alone. We discuss how learned behavioral models can support explainability, model-guided exploration, auditing, and analysis of agentic systems. We further demonstrate symbolic model-based knowledge transfer from powerful frontier models to compact language models. In addition, we show how model transformations can derive concise explanations of agent behavior in a penetration-testing case study comprising 12 vulnerable machines. ATLAS highlights a new opportunity for model-driven engineering: transforming agent trajectories into explicit behavioral models that enable systematic understanding and analysis of otherwise opaque AI agents.
\end{abstract}

%%
%% http://dl.acm.org/ccs.cfm.



\maketitle

\begin{center}
\small
This is the author's version of the work. The definitive version will be
published in the ACM/IEEE 29th International Conference on Model Driven
Engineering Languages and Systems (MODELS 2026).
\url{https://doi.org/10.1145/3822455.3838777}
\end{center}
\section{Introduction}
Agentic AI systems~\cite{ReAct} are increasingly being used to solve tasks that require \emph{many sequential steps}. These tasks are becoming increasingly complex: agents must often invoke external tools~\cite{DBLP/corr/abs-2302-04761}, maintain and retrieve information from memory, interact with external environments, incorporate human or environmental feedback, and adapt their behavior over time. Despite these challenges, agentic AI systems have demonstrated remarkable success across a range of domains, including realistic web navigation, software engineering, robotic coordination, and medical decision support~\cite{DBLP/air/AliDC26}. 

Current approaches record trajectories of agent-environment interactions or construct models of the environment to steer agentic behavior~\cite{li2026preactcomputerusingagentsfaster}.  However, there is a notable gap in research on constructing explicit behavioral models of the agent itself. Without such models, model-based analysis, comparison, monitoring, and validation remain difficult, but the complexity of LLM-driven decision-making renders manual modeling practically infeasible. We argue that recorded trajectories are an untapped source of information about agent behavior. We therefore propose to automatically recover \textbf{compact behavioral models of AI agents through automata learning}, to capture high-level task-solving strategies.

\paragraph{\textbf{A New Direction for Modeling Agentic AI: Behavioral Models via Automaton Abstractions}}
In this work, we recover behavioral models of agentic AI systems from observed interaction traces. In particular, we use labelled Markov chains (MCs), which provide a natural representation for sequential stochastic processes: they describe temporal and stochastic behavior and capture uncertainty, as required by stateful agents backed by foundation models. 
Labelled MCs provide compact state-based models of sequential behavior. They expose recurring interaction patterns, branching decisions, loops, and stochastic choices while supporting automated analyses such as probabilistic model checking~\cite{storm,baier2018model}.

We propose \textbf{ATLAS} (Automata Learning for Agent Trajectory Analysis and Strategy Discovery). This approach automatically learns MCs that serve as behavioral models of AI decision-making for solving a given task. The proposed methodology includes the following steps.
\begin{description}
    \item[Trace collection.] We collect interaction traces produced by the agent while solving a given task.
    \item[Trace abstraction.] Since raw actions and observations can be highly specific, noisy, and semantically redundant, we leverage foundation models to compute \emph{abstractions} that group similar agent actions and environmental observations into \emph{higher-level categories}.  
    \item[Automata learning.] We apply \emph{automata learning}~\cite{higuera,DBLP:journals/cacm/Vaandrager17} to infer MCs from abstracted traces with states labeled by abstract action-observation categories. 
    \item [Model-Driven Engineering.] The learned models become first-class engineering artifacts that support downstream model-driven engineering tasks such as explanation, runtime monitoring~\cite{wang2026probguardprobabilisticruntimemonitoring,DBLP:journals/corr/abs-2503-18666}, auditing, model slicing, symbolic knowledge transfer, and formal analysis.
\end{description}
Foundation models provide semantic abstraction, whereas automata learning reconstructs the underlying behavioral structure shared across executions. By learning states and transitions based on differences in observable behavior, automata learning can infer latent states of agentic decision-making.

This paper makes the following contributions:
\begin{itemize}
    \item an LLM-guided abstraction pipeline for agent trajectories, 
    \item an automata-learning approach for recovering behavioral models,
    \item a proof-of-concept evaluation on penetration-testing agents, demonstrating two applications of the recovered models (explanation and symbolic knowledge transfer).
\end{itemize}

\paragraph{\textbf{Related work}} Current approaches to making agentic AI systems more transparent include methods such as chain-of-thought-style reasoning~\cite{Wei0SBIXCLZ22}, interleaved reasoning-action traces~\cite{ReAct}, self-reflection~\cite{ShinnCGNY23}, task decomposition~\cite{YaoYZS00N23}, and process-level evaluation~\cite{LightmanKBEBLLS24}. AgentGuard~\cite{DBLP:conf/kbse/Koohestani25} constructs stochastic models for verification, but relies on manual abstraction. ReAct-style approaches~\cite{ReAct} interleave generated reasoning steps with external actions, such as searching, querying tools, or interacting with an environment. This idea is closely related to chain-of-thought explanations, in which models expose intermediate reasoning steps to make their outputs easier to inspect~\cite{Wei0SBIXCLZ22}. \emph{However, such traces primarily describe a particular execution trajectory; they do not necessarily explain the agent's broader task-solving strategy}. Closest in spirit to our work are \textsc{ProbGuard}~\cite{wang2026probguardprobabilisticruntimemonitoring} and Trace2Chain~\cite{DBLP:journals/corr/abs-2604-24579}, both of which create Markov chains for runtime monitoring and measuring the reliability of LLM agents, respectively. Trace2Chain and \textsc{ProbGuard} fit Markov chains over explicit discrete state representations (which may satisfy the Markov property), whereas our automata-learning approach infers latent behavioral states directly from abstract interaction traces. Preact~\cite{li2026preactcomputerusingagentsfaster} constructs finite-state models of the environment to make repeated agentic tasks more efficient.

Closely related to our work, Ayoughi et al.~\cite{DBLP:conf/models/AyoughiNSS24} combine passive automata learning with machine learning-based abstraction in a security case study. They employ decision trees for abstraction and learn Moore machine models from an intrusion detection system. In contrast, we use LLM-guided abstraction combined with stochastic automata learning to learn behavioral models of agentic AI systems.

Recent work has explored
LLM-based abstraction for trajectory understanding in various contexts, e.g.,
Koohestani et al.~\cite{koohestani2026tricegar} use decision-tree predicates
to construct abstract state spaces from agent execution logs for MDP-based
runtime verification; Sui et al.~\cite{sui2026thinklessknowmore} cluster chain-of-thought reasoning steps into latent Markov-chain states. Recent work has also explored extracting knowledge from agent trajectories in natural-language form: TIM~\cite{DBLP:journals/corr/abs-2603-10600} extracts different types of \emph{tips}, Trace2Skill~\cite{DBLP:journals/corr/abs-2603-25158} creates so-called \emph{skills} from trajectories, and ExpeL~\cite{DBLP:conf/aaai/Zhao0XLLH24} extracts \emph{insights}. These artifacts are designed to guide future LLM agents rather than to serve as formal behavioral models amenable to automated analysis. Nevertheless, the recency of these publications suggests growing interest in learning reusable knowledge from agent trajectories.

To the best of our knowledge, no existing approach combines semantic LLM-based abstraction with automata learning to recover behavioral models of AI agents from execution trajectories.

\textbf{Outline.}
Throughout, we use agentic penetration testing to illustrate the steps of ATLAS and demonstrate its usefulness. We have chosen this application domain as it is a high-stakes domain of particular societal relevance, highlighted by recent concerns regarding the cyber-offensive capabilities of frontier language models.~\cite{aisi2025claude}. 

Section~\ref{sec:example} presents an illustrative example providing an overview of our approach applied to penetration testing. It exemplifies model-based analyses enabled by learned MCs. Section~\ref{sec:method} introduces the proposed model recovery pipeline in more detail, including the LLM-enabled abstraction of actions and observations, and the learning of MCs. Section~\ref{sec:experiments} reports on initial experimental results, demonstrating the potential of the approach by deriving compact MC abstractions from benchmark penetration-testing scenarios and applying them in two use cases. We conclude the paper in Sect.~\ref{sec:concl} with a summary of our findings and provide an outlook on future plans in Sect.~\ref{sec:future}.

\section{Learning a Penetration Testing Strategy}
\label{sec:example}
We demonstrate ATLAS by recovering symbolic behavioral models of a penetration-testing AI agent. In particular, we execute \emph{HackingBuddyGPT}~\cite{Happe_2023} in combination with different LLMs on a benchmark suite of vulnerable machines~\cite{happe2024got}. Given recorded trajectories, we learn models that capture the agent's behavioral strategy during penetration testing and the system response.

\definecolor{agentbg}{RGB}{220,235,255}
\definecolor{vmbg}{RGB}{220,255,220}
\definecolor{catbg}{RGB}{255,255,220}
\definecolor{autbg}{RGB}{255,235,220}
\definecolor{successbg}{RGB}{200,255,200}
\definecolor{errbg}{RGB}{255,200,200}

\begin{figure*}[t]
\centering
% ============================================================================
% Panel (a): Concrete Interaction Sequence Diagram
% ============================================================================
\fboxsep=2pt
\setlength{\fboxrule}{0.3pt}

\fcolorbox{gray!50}{white}{
\begin{minipage}[t][4.1cm][t]{0.26\textwidth}
\centering
\footnotesize\textbf{(a) Concrete Interaction}\\
\begin{tikzpicture}[
    lifeline/.style={draw, thick},
    ahead/.style={draw, fill=agentbg, minimum width=1.1cm, minimum height=0.5cm, font=\scriptsize\bfseries},
    vhead/.style={draw, fill=vmbg, minimum width=1.1cm, minimum height=0.5cm, font=\scriptsize\bfseries},
    msg/.style={->, >=stealth, font=\tiny\tt, align=left},
    ret/.style={->, >=stealth, dashed, font=\tiny, align=left},
]
\node[ahead] (A) at (0,0) {Agent};
\node[vhead] (V) at (3.5,0) {VM (SSH)};
\draw[lifeline] (A) -- (0,-3.6);
\draw[lifeline] (V) -- (3.5,-3.6);

% Step 1: id
\draw[msg] (0,-0.5) -- (3.48,-0.5) node[midway,above,font=\tiny\tt] {exec\_command id};
\draw[ret] (3.5,-0.65) -- (0.02,-0.65) node[midway,below,font=\tiny] {uid=1000(lowpriv) gid=1000(lowpriv)\ldots};

% Step 2: sudo -l
\draw[msg] (0,-1.25) -- (3.48,-1.25) node[midway,above,font=\tiny\tt] {exec\_command sudo -l};
\draw[ret] (3.5,-1.4) -- (0.02,-1.4) node[midway,below,font=\tiny] {Matching Defaults entries for lowpriv\ldots\\ (ALL) NOPASSWD:ALL\ldots};

% Step 3: sudo id (optional verification)
\draw[msg] (0,-2.2) -- (3.48,-2.2) node[midway,above,font=\tiny\tt] {exec\_command sudo id};
\draw[ret] (3.5,-2.35) -- (0.02,-2.35) node[midway,below,font=\tiny] {uid=0(root) gid=0(root) groups=0(root)};

% Step 4: sudo -i
\draw[msg] (0,-2.95) -- (3.48,-2.95) node[midway,above,font=\tiny\tt] {exec\_command sudo -i};
\draw[ret] (3.5,-3.1) -- (0.02,-3.1) node[midway,below,font=\tiny] {root@af467fadbde3:\string~{}\#};

\node[font=\small\bfseries, fill=successbg] at (1.75,-3.5) {Got root!};
\end{tikzpicture}
\end{minipage}
}
\hfill
% ============================================================================
% Panel (b): Abstract Interaction Diagram
% ============================================================================
\fcolorbox{gray!50}{white}{
\begin{minipage}[t][4.1cm][t]{0.24\textwidth}
\centering
\footnotesize\textbf{(b) Abstracted Interaction}\\
\begin{tikzpicture}[
    lifeline/.style={draw, thick},
    ahead/.style={draw, fill=agentbg, minimum width=1.0cm, minimum height=0.5cm, font=\scriptsize\bfseries},
    vhead/.style={draw, fill=catbg, minimum width=1.0cm, minimum height=0.5cm, font=\scriptsize\bfseries},
    msg/.style={->, >=stealth, font=\footnotesize, align=left},
    ret/.style={->, >=stealth, dashed, font=\footnotesize, align=left},
]
\node[ahead] (A) at (0,0) {Agent};
\node[vhead] (V) at (3.0,0) {VM};
\draw[lifeline] (A) -- (0,-3.6);
\draw[lifeline] (V) -- (3,-3.6);

% Step 1
\draw[msg] (0,-0.5) -- (2.98,-0.5) node[midway,above=-0.1cm,font=\footnotesize] {cmd.exec (id)};
\draw[ret] (3,-0.65) -- (0.02,-0.65) node[midway,below,font=\footnotesize] {DISCOVERY\_USER};

% Step 2
\draw[msg] (0,-1.25) -- (2.98,-1.25) node[midway,above=-0.1cm,font=\footnotesize] {cmd.exec (sudo -l)};
\draw[ret] (3,-1.4) -- (0.02,-1.4) node[midway,below,font=\footnotesize] {EXPLOITABLE\_SUDO};

% Step 3
\draw[msg] (0,-2.2) -- (2.98,-2.2) node[midway,above=-0.1cm,font=\footnotesize] {cmd.exec ( id)};
\draw[ret] (3,-2.35) -- (0.02,-2.35) node[midway,below,font=\footnotesize] {EXPLOITABLE\_ROOT};

% Step 4
\draw[msg] (0,-2.95) -- (2.98,-2.95) node[midway,above=-0.1cm,font=\footnotesize] {priv.esc (sudo -i)};
\draw[ret] (3,-3.1) -- (0.02,-3.1) node[midway,below,font=\footnotesize] {EXPLOITABLE\_ROOT\_\_success};
\end{tikzpicture}
\end{minipage}
}
\hfill
% ============================================================================
% Panel (c): Automaton (hb_mc_tiny.dot) — tikz automata library
% ============================================================================
\fcolorbox{gray!50}{white}{
\begin{minipage}[t][4.1cm][t]{0.41\textwidth}
\centering
\footnotesize\textbf{(c) Learned Markov Chain}\\
\begin{tikzpicture}[
    ->, >=stealth, auto, node distance=1.6cm, semithick,
    every state/.style={fill=autbg, draw=black, thick,
                        align=center, font=\tiny,
                        inner sep=0pt},
    acc/.style={accepting, fill=successbg},
    err/.style={fill=errbg},
    other/.style={fill=gray!15,dashed},
    font=\scriptsize,
]

\node[state, initial,initial where=right,
      label={[xshift=1mm,yshift=-2mm]above right:{init}}] (q0) {$q_0$};

\node[state,
      label=below:{\shortstack{cmd.exec (id)\\DISCOVERY\_USER}}]
      (q1) [below left=0.3cm and 0.3cm of q0] {$q_1$};

\node[state,
      label=below:{\shortstack{enum.user\\DISCOVERY\_USER}}]
      (q4) [below left=-0.4cm and 1.8cm of q0] {$q_3$};

\node[state, err,
      label=below:{\shortstack{cmd.exec (malformed)\\ERROR\_NONE}}]
      (q5) [below=1.3cm of q4] {$q_4$};

\node[state,
      label=below:{\shortstack{cmd.exec (sudo -l)\\EXPLOITABLE\_SUDO}}]
      (q2) [right=of q1] {$q_2$};

\node[state, acc,
      label=below:{\shortstack{priv.esc(sudo -i)\\EXPLOITABLE\_ROOT\_\_success}}]
      (q9) [below left=1.2cm and 0.6cm of q2] {$q_8$};
  \node[state,
      label=below:{\shortstack{cmd.exec(id)\\EXPLOITABLE\_ROOT}}]
      (q7) [below right=1.0cm and 0.8cm of q2] {$q_7$};

\node[state, other,
      label=below:{other}]
      (q6) [right=0.7cm of q2] {};

\path (q0) edge node[above left,font=\scriptsize] {0.60} (q1)
           edge node[above=0.05cm,font=\scriptsize] {0.35} (q2)
           edge node[above,font=\scriptsize] {0.05} (q4)
      (q1) edge node[above,font=\scriptsize] {0.70} (q2)
           edge node[below left=-0.2cm and 0.1cm,font=\scriptsize] {0.30} (q5)
      (q4) edge node[left,font=\scriptsize] {1.00} (q5)
      (q5) edge node[below right=-0.6cm and 0.5cm,font=\scriptsize] {1.00} (q2)
      (q2) edge node[below left =0.1cm and 0.2cm,font=\scriptsize] {0.30} (q9)
           edge[dashed] node[above,font=\scriptsize] {0.60} (q6)
           (q2)
           edge node[below right=0.1cm and 0.3cm,font=\scriptsize] {0.10} (q7)
           (q7)
           edge node[above,font=\scriptsize] {1.00} (q9)
           (q9) edge[loop left] node[font=\scriptsize] {1.00}(q9);
\end{tikzpicture}
\end{minipage}
}

\caption{ATLAS automata learning pipeline applied on HackingBuddyGPT~\cite{Happe_2023} executing Scenario 3 of~\cite{happe2024got} (\texttt{sudo NOPASSWD:ALL} privilege escalation).
Panel~(a): concrete Agent--VM interaction trace comprising 4~steps.
Panel~(b): LLM-abstracted categorical trace shown as an abstract interaction.
Panel~(c): simplified Markov chain learned from 20 abstract traces, with edge labels showing empirical transition probabilities.}
\label{fig:illustrative_example}
\Description{Illustrative example showing the application of ATLAS. Left panel: concrete agent-environment interaction; middle panel: abstracted agent-environment interaction; Right panel: A learned Markov chain for this scenario.}
\end{figure*}
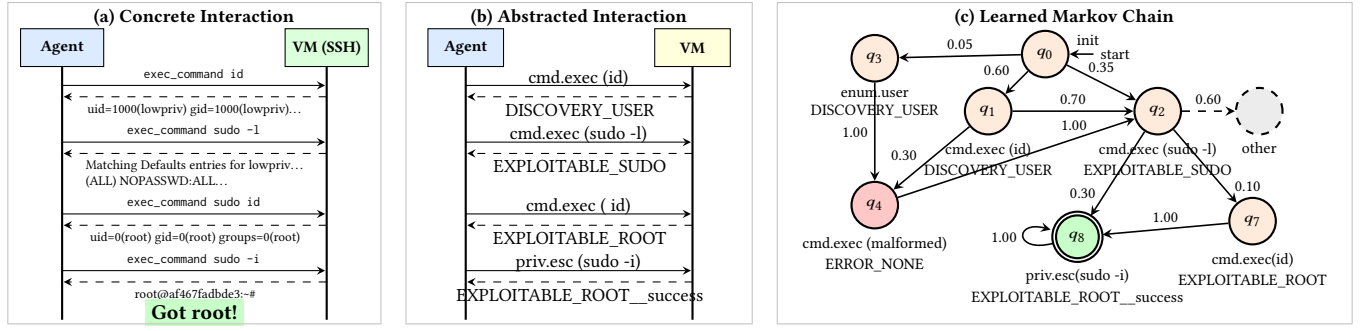

 In the following, we illustrate our learning approach on Docker Scenario 3 from~\cite{happe2024got}.
Figure~\ref{fig:illustrative_example} shows how we get from automated agentic penetration testing in this scenario to a concise symbolic model of the observed behavior. The scenario includes a simple privilege escalation vulnerability: a Linux host exposing a misconfigured sudo policy that grants a low-privileged user passwordless execution of arbitrary commands as root.

\noindent
\textbf{Step 1: Penetration Testing.}
The left panel depicts the interaction between the agent and the vulnerable host, i.e., its environment:
\begin{itemize}
    \item \textbf{Agent} $\rightarrow$ \textbf{Environment}: Semi-structured actions, primarily bash commands executed via SSH.
    \item \textbf{Environment} $\rightarrow$ \textbf{Agent}: Unstructured observations as com\-mand-line output in response to issued commands.
\end{itemize}
We can see that the agent is successful, i.e., it opened a root shell, and a security analyst can readily understand why. The crucial command is \texttt{sudo -l} and the corresponding observation, which has been truncated, that tells us that we have privileges to run an arbitrary command. What observing this interaction does not tell us:
(1) Does this \emph{always} work? (2) Which observations influence the agent's decisions?
(3) Why does the agent execute \texttt{sudo id} despite already having sufficient information to escalate privileges?
    
\noindent
\textbf{Step 2: Trajectory Abstraction.}
Looking at the concrete interaction, we can see that there are already quite a few details, even though it is a very simple example. Learning a behavioral model directly from concrete interactions is likely infeasible for all but the most trivial examples. Therefore, we perform an abstraction over pairs of actions performed by the agent and observations produced by the environment.

The middle panel shows the abstracted interaction corresponding to the concrete interaction. A three-phase LLM-based process automatically creates such abstract sequences, with the goal of succinctness and usefulness. The abstraction process maps
\begin{itemize}
    \item each \textbf{action} to an \textbf{action category} followed by an optional action target in parentheses, and 
    \item each \textbf{observation} to an observation category describing both the usefulness of the output and the affected system entity.
\end{itemize}

The abstraction intentionally sacrifices low-level detail in favor of recurring behavioral patterns: some steps are already easier to understand. We can observe that \texttt{sudo -l} leads to a discovery of something exploitable. 

\noindent
\textbf{Step 3: Automata Learning.}
The third step applies Alergia, an automata learning algorithm~\cite{DBLP:conf/icgi/CarrascoO94} adapted to MCs~\cite{DBLP:journals/ml/MaoCJNLN16}, over abstract interaction sequences to learn a Markov chain. In the right panel, we can see a Markov chain learned from $20$ abstracted interaction sequences, including the sequences from the middle panel. 

The states of the Markov chain are labeled with pairs of abstract actions and observations plus an optional \emph{success} label, and the edges are labeled with probabilities. It summarizes the agent's behavioral strategy together with the corresponding system responses, and reveals several patterns:
\begin{enumerate}
\item The agent usually begins with reconnaissance by querying user information or checking available sudo privileges via \texttt{sudo -l}.
\item After user discovery, the agent either:
    \begin{itemize}
        \item produces a malformed command, a recurrent failure mode of the LLM, or
        \item proceeds to \texttt{sudo -l}, which reveals exploitable sudo privileges.
    \end{itemize}
\item Upon observing \texttt{EXPLOITABLE\_SUDO}, the agent succeeds immediately in 30\% of the cases by executing \texttt{sudo -i}. In the remaining cases, it performs additional exploratory actions before eventually escalating privileges.
\end{enumerate}

At this point, we can revisit the questions from above: (1) The learned model suggests that all observed execution paths eventually reach a successful privilege escalation (also in the omitted \emph{other} part). (2) \texttt{sudo -l} leads to success within two steps with $0.4$ probability; thus, the response to this command is likely important. Observing user information often leads to malformed commands, which we should investigate. (3) We do not know exactly why the agent performs \texttt{sudo id}, although it should know that an immediate exploit is possible, but we know that it makes inefficient choices more often than not (70\% of the cases). In Sect.~\ref{sec:experiments}, we demonstrate how the recovered behavioral model supports multiple downstream analyses. In particular, we show how the symbolic knowledge encoded in the learned Markov chain can be transferred to other LLMs and use model slicing to derive concise explanation models.
\section{LLM-Guided Abstraction \& Model Learning}
\label{sec:method}
This section presents the two main components of ATLAS: the \textbf{abstraction process}, which translates raw interaction traces into a compact symbolic alphabet, and the \textbf{model learning} stage, which uses this alphabet to infer behavioral models as probabilistic finite-state models.

\textbf{Abstraction. }
A central challenge in learning behavioral models from agent trajectories is
the \emph{vocabulary gap}: raw traces consist of concrete shell commands, tool invocations, and unstructured textual outputs. The goal of abstraction is to replace low-level commands and observations with semantically meaningful symbols that preserve behavioral intent while enabling automata learning.

\begin{figure}

\resizebox{\columnwidth}{!}{
\begin{tikzpicture}[
    block/.style={rectangle, draw, minimum width=1.8cm, minimum height=1.2cm, 
                  align=center, rounded corners, fill=blue!10, text width=1.7cm,
                  font=\sffamily\small},
    io/.style={cylinder, shape border rotate=90, draw, minimum width=1.7cm, 
               minimum height=2.0cm, align=center, fill=red!10, text width=2.5cm, aspect=0.3,
               font=\sffamily\small},
    subblock/.style={rectangle, draw, minimum width=3.3cm, minimum height=2.8cm, 
                     align=center, rounded corners, fill=green!10, 
                     font=\sffamily\small, text width=2.8cm},
    detailblock/.style={rectangle, draw, minimum width=3cm, minimum height=0.6cm,
                        align=center, rounded corners, fill=white, font=\sffamily\small},
    abstractionbox/.style={draw, dashed, thick, inner sep=0.8cm, fill=gray!5,
                           rounded corners},
    arrow/.style={->, >=stealth, line width=1.5pt},
    feedback/.style={->, >=stealth, line width=1.5pt, dashed, bend right=25},
    label/.style={font=\sffamily\bfseries\small}
]

% ---- Input ----
\node [io] (input) at (0,-.35) {Agent Trajectories\\ (raw logs)};

% Category Discovery block
\node [subblock] (discovery) at (3.5,0) {};
\node [label, above=-.1cm of discovery] (discoverylabel) {Category Discovery};
% Inner details inside discovery
\node [detailblock, below=0.1cm of discovery.north, anchor=north] (batch) {Batch interactions};
\node [detailblock, below=0.4cm of batch] (propose) {(1) LLM categorization};
\node [detailblock, below=0.4cm of propose] (normalize) {(2) Category normalization};

% Arrows inside discovery
\draw [arrow] (batch.south) -- (propose.north);
\draw [arrow] (propose.south) -- (normalize.north);

% Per-Trace Assignment block
\node [subblock, right=.4cm of discovery] (assignment) {};
\node [label, above=-.1cm of assignment] (assigmentlabel) {Trajectory Abstraction};
\node [detailblock, below=0.1cm of assignment.north, anchor=north] (perrun) {(3) Per-trace assignment};
\node [detailblock, below=0.4cm of perrun] (struct) {Structured format};
\node [detailblock, below=0.4cm of struct] (retry) {Retry mechanism};

\draw [arrow] (perrun.south) -- (struct.north);
\draw [arrow] (struct.south) -- (retry.north);

% Connect the two main blocks
\draw [arrow] (discovery.east) -- (assignment.west);

% ---- Draw dashed box around abstraction ----
% \node [abstractionbox, fit=(discovery) (assignment)] (abstraction) {};
% \node [label, above] at (abstraction.north) {LLM-Guided Abstraction};

% ---- Output after abstraction ----
\node [block, right=.4cm of assignment] (abstract) {Abstract\\ Traces};

% ---- Main data flow arrows ----
\draw [arrow] (input.14) -- (discovery.west);
% arrow from abstraction to abstract goes from assignment.east to abstract.west
\draw [arrow] (assignment.east) -- (abstract.west);

% ---- Feedback loop ----
% \draw [feedback] (abstract.south) to[out=-90,in=-90,looseness=1.8] 
%     node[midway, below, align=center, font=\sffamily\footnotesize] 
%     {Model Refinement\\ (adjust abstraction)}
%     (discovery.south);

\end{tikzpicture}
}

    \caption{LLM-Guided abstraction}
    \label{fig:llm_guided_abstraction}
\Description{An illustration of LLM-guided abstraction of concrete agent-environment interactions.}
\end{figure}
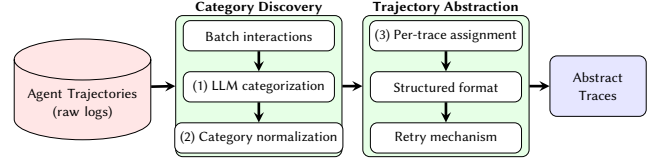

Given $k$ environment-agent interaction traces, our approach to trace abstraction is \emph{fully LLM-driven} and operates in three stages (see Fig.~\ref{fig:llm_guided_abstraction}):

\begin{enumerate}
  \item \textbf{LLM categorization.} Interactions from all $k$ runs are bat\-ched and presented to the  LLM with a structured prompt, which instructs it to map:
  \begin{itemize}
      \item \textbf{concrete actions} to 
      \textbf{action categories}, optionally followed by an action target in parentheses, and map
      \item \textbf{concrete observations} to \textbf{observation categories} composed using a two-component \textsc{Value\_Target} format, e.g.,  \texttt{DISCOVERY\_SUID}, \texttt{EXPLOITABLE\_FILE}, and \texttt{ERROR\_NONE}.
  \end{itemize}
    The LLM proposes action categories, which describe \emph{what} the agent did, and output categories, which describe the \emph{semantic value and the target} of the observation resulting from the action
    \item \textbf{Category normalization.} Category labels are accumulated across runs, creating redundant categories describing semantically equivalent actions and observations.

    To produce a compact set of labels, we provide all discovered categories to the  LLM with a prompt to normalize them by merging semantically similar labels.

  \item \textbf{Per-trace assignment.} For each run, we present the full interaction sequence and normalized category vocabulary to the LLM, which maps each interaction to an (action-category, output-category) pair in a structured format, producing an abstract trace suitable for automata
    learning. A retry mechanism with format correction prompts handles
    occasional malformed LLM outputs.
\end{enumerate}

The abstraction step bridges between unstructured agent behavior
and formal model inference. Unlike approaches that use coarse pre-defined manual abstraction functions or manual feature engineering, our LLM-driven categorization adapts
to the semantic content of each interaction in each environment and task to produce a stable symbolic
alphabet. Automata learning uses the alphabet to recover explicit behavioral models.

\textbf{Model learning.}
To learn behavioral models, we propose applying Alergia~\cite{DBLP:journals/ml/MaoCJNLN16}, a state-merging-based automata learning algorithm. The algorithm takes abstracted interaction sequences as inputs, arranges them in a tree, and creates abstract Markov chains (MCs) through iterated tree node merging followed by transition probability normalization. The resulting MCs are probabilistic models over abstract action-observation pairs, capturing the agent's behavior and the environment's response. 

\noindent\textbf{Properties of ATLAS.}
% \begin{compactitem}
% \item 

\noindent\textbf{Fully Automated. }
Raw interaction trajectories are transformed into concise behavioral models with little parameterization required.
% \item 

\noindent\textbf{State Detection.} Alergia detects states based on observational differences in future behavior. Consequently, it infers latent states in the agent's decision-making without relying on state-extraction heuristics from the environment~\cite{li2026preactcomputerusingagentsfaster}. Unlike existing approaches, it does not require Markovian abstract observations~\cite{DBLP:conf/kbse/Koohestani25,DBLP:journals/corr/abs-2604-24579}, but can infer latent Markov states that capture relevant history from partial observations through automata learning~\cite{DBLP:conf/ifm/MuskardinTAP23}.
    % \item 
    
    \noindent\textbf{Cost of Abstraction. } Category discovery requires a powerful LLM, as abstract categories need to be identified from scratch. Abstraction assignment is a classification task, which can be performed by smaller language models (as illustrated in the dynamic mode in Sect.~\ref{sec:experiments}). Consequently, the abstraction of additional traces for model refinement on additional data is relatively cheap.
    % \item 
    
    \noindent\textbf{Generality.} Our focus is on behavioral models of agents, but ATLAS naturally extends to other modelling formalisms to support additional uses. Learning of \textbf{discrete Markov decision models} via IoAlergia~\cite{DBLP:journals/ml/MaoCJNLN16} can provide an environ\-ment-centric view while capturing stochastic behavior of low-level decision-making. Learning of non-deterministic (and non-stochastic) models through k-tails~\cite{biermann,gsm} can provide a view focused on environment behavior.
% \end{compactitem}

\section{Case Study: Applications of Learned Behavioral Models}
\label{sec:experiments}
In this section, we report on a proof-of-concept application of ATLAS, learning Markov chain models of HackingBuddyGPT~\cite{Happe_2023} on 12 vulnerable Linux virtual machines from~\cite{happe2024got}\footnote{The experiments exclude Scenario 4 of the thirteen, as interactive shell commands, which the SSH-based setup does not support}. We demonstrate two applications of the recovered models: transferring symbolic task knowledge from frontier to compact language models, and deriving concise explanations of successful behavioral strategies.

\paragraph{ATLAS Setup}
For trace collection, we performed $k=20$ penetration testing runs with DeepSeek V4 Flash (non-reasoning mode, 128K~context) on each VM with a 20-turn limit. We use the same model to generate abstract interaction traces via the approach from Fig.~\ref{fig:llm_guided_abstraction}.
Additionally, we augment the final observation in a run with \texttt{success}/\texttt{fail} to indicate if the run was successful.
Finally, we apply Alergia implemented in AALpy~\cite{DBLP:journals/isse/MuskardinAPPT22} to learn a labelled MC for each scenario from the $k$ abstracted traces. Alongside each learned behavioral model and abstraction, we store concretization mappings, which record frequent concrete commands for each abstract action category. These can be used to ground abstract actions.

\subsection{Knowledge Transfer from Learned Markov Chains}
We explore four modes of 
extracting transferable knowledge from learned MCs with varying LLM dependence.

We call the extracted knowledge \emph{strategy summaries}, which we use to steer agentic penetration tests with a weaker, cheaper \emph{executor} LLM (ministral-14b
or ministral-8b, both with 262K context), by augmenting the system prompt. While we accessed these LLMs via \url{openrouter.ai}, they can be run on consumer GPUs. To examine transfer viability, we perform 5 penetration tests of up to 20 turns on each VM with each transfer mode and compare to an unaided baseline to establish the lower bound of executor capability. The full pipeline is implemented in approximately 3400 lines of Python, integrated with the HackingBuddyGPT~\cite{Happe_2023} framework. The learned MCs, the abstract traces, the concrete traces, and the source code are included in the replication package available at \url{https://doi.org/10.6084/m9.figshare.32841767}.

\subsubsection{Transfer Modes}
The first two transfer modes query a frontier model (DeepSeek V4 Pro), which we call \emph{knowledge LLM} once per VM to generate a \emph{strategy summary}. 

In the \textbf{minimal} mode, the knowledge LLM receives a compact summary of the learned MCs and executed commands. It is prompted to produce a concise strategy of 2--3 steps.

\textbf{Guided} mode differs from \emph{minimal} in that the knowledge LLM produces a structured, more verbose guide, including a general strategy, up to 5 concrete steps, and exploration suggestions.

\textbf{Common Path} mode analyzes the learned MC algorithmically, producing strategy summaries comprised of the most common steps leading to success.

\textbf{Dynamic} mode provides adaptive guidance \emph{during} the
penetration test. To this end, we traverse learned MCs in parallel to the environments. At each step, (1) we determine a promising action and suggest this single action to the executor LLM. (2) After each concrete command execution by the executor, a cheap classifier LLM -- we simply use the executor LLM -- maps the concrete (command, observation) pair to known action and observation categories. (3) Based on this abstract label, we advance in the learned MC. Dynamic strategy summaries are formatted as compact one-liners comprising feedback on the last step, the suggested next command, and an exploration reminder.

Dynamic mode includes an anti-repetition mechanism and dead-end detection. It separates: 
\begin{itemize}
    \item \emph{Symbolic high-level planning} driven by the recovered behavioral model, and 
    \item \emph{low-level execution} performed by the executor LLM.
\end{itemize}

\subsubsection{Results}

\begin{table}[t]
\centering
\caption{Transfer mode comparison for ministral-8b.}
\label{tab:transferministral8b}
\small
\begin{tabular}{lccc}
\toprule
Transfer Mode & Roots/Runs & Success Rate & VMs Exploited \\
\midrule
Baseline (no hints)  & 3/60 & 5.0\% & 3, 6 \\
Common path          & 1/60 & 1.7\% & 13 \\
Minimal              & 5/60 & 8.3\% & 13 \\
Guided               & 5/60 & 8.3\% & 13 \\
Dynamic              & 17/60 & 28.3\% & 3, 7, 8, 13 \\
\bottomrule
\end{tabular}
\end{table}

\begin{table}[t]
\centering
\caption{Transfer mode comparison for ministral-14b.}
\label{tab:transferministral14b}
\small
\begin{tabular}{lccc}
\toprule
Transfer Mode & Roots/Runs & Success Rate & VMs Exploited \\
\midrule
Baseline (no hints)  & 1/60 & 1.7\% & 5 \\
Common path          & 7/60 & 11.7\% & 2, 13 \\
Minimal              & 7/60 & 11.7\% & 5, 13 \\
Guided               & 13/60 & 21.7\% & 1, 2, 8, 13 \\
Dynamic              & 23/60 & 38.3\% & 2, 3, 5, 7, 8, 13 \\
\bottomrule
\end{tabular}
\end{table}

Tables~\ref{tab:transferministral8b} and \ref{tab:transferministral14b} show results of an experimental evaluation of the different transfer modes and baseline mode (unguided), using ministral-8b and ministral-14b as executor LLMs. Table~\ref{tab:transferministral14b} shows that the hints provide sufficient information to replicate successful behavior, albeit not in all scenarios. While the success rate generally increases, we further observe that the transfer mode has a large impact. The dynamic mode is clearly the most effective, both in the success rate and unique scenario successes. This is in line with observations we made during experiments: small models fail to follow multi-step instructions as they struggle to keep track of the current state. 
Dynamic mode alleviates this issue by tracking an abstract state -- the current state in the learned MC -- and suggesting only a single step. It constitutes a successful integration of model-driven planning and AI-based execution.

The same pattern can be observed for the 8b model in Table~\ref{tab:transferministral8b}, where the \emph{dynamic} mode is the most effective. Additionally, we can see that the \emph{common path} decreased the success rate. This mode creates relatively large hints that may be hard to process for small models. While we do not achieve replicating all successful behavior, the results suggest that symbolic knowledge transfer is possible. The recovered behavioral models enabled successful symbolic transfer from a frontier model  (DeepSeek-V4-Flash) with 284B parameters to a 14B parameter model (ministral-14b), which solved 50\% of the benchmark tasks despite its compact size.

\subsection{Explanation Model Extraction}
Penetration testing agents typically start with a reconnaissance phase, captured by learned strategy models. As a result, learned MCs include details that are irrelevant to task completion, making interpretation difficult. Similar behavior affects other application domains, e.g., source code exploration helps coding agents localize relevant code before implementing changes~\cite{zhang2026sweexplorebenchmarkingcodingagents}, but analyzing exploration activities likely provides limited insights. 
To address this issue, we propose \textbf{observation-focused extraction of explanation models} from learned MCs, succinctly capturing task-relevant behavior. Below, we illustrate two methods for explaining behavior leading to the observation label $ol=\texttt{success}$.

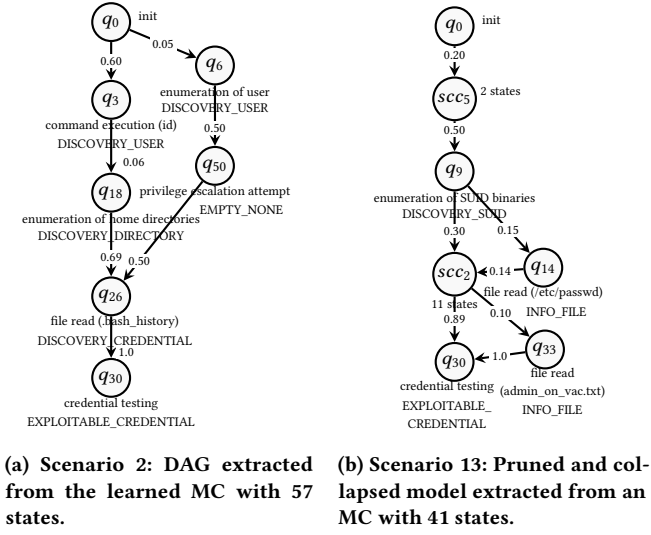
\begin{figure}[t]
    \centering
    \begin{subfigure}[t]{0.48\columnwidth}
        \centering
        \begin{tikzpicture}[
  ->,
  >=stealth,
  node distance=1.0cm and 0.8cm,
  every state/.style={draw, rounded corners, thick, fill=gray!5,
                       font=\footnotesize\tt, minimum size=5mm,
                       inner sep=1pt},
  every edge/.style={draw, ->, >=stealth, thick},
  edge label/.style={font=\tiny, fill=white, inner sep=1pt},
  desc/.style={font=\tiny, align=center,yshift=-1mm,yshift=-1mm},
]
\node[state] (q0) {$q_0$};
\node[desc,xshift=0.3cm,yshift=0.1cm] at (q0.north east) {init};

  \node[state, below = 0.5cm of q0] (q3) {$q_3$};
  \node[desc] at (q3.south) {\shortstack{command execution (id)\\DISCOVERY\_USER}};
  
  \node[state,below=0.7cm of q3] (q18) {$q_{18}$};
  \node[desc] at (q18.south) {\shortstack{enumeration of home directories\\DISCOVERY\_DIRECTORY}};

  \node[state,below=3.1cm of q0] (q26) {$q_{26}$};
  \node[desc,xshift=0.5mm] at (q26.south) {\shortstack{file read (.bash\_history)\\DISCOVERY\_CREDENTIAL}};
  
  \node[state,below = 0.55cm of q26] (q30) {$q_{30}$};
  \node[desc] at (q30.south) {\shortstack{credential testing\\EXPLOITABLE\_CREDENTIAL}};
  
  \node[state,below right= 0.2cm and 1cm of q0] (q6) {$q_6$};
  \node[desc] at (q6.south) {\shortstack{enumeration of user\\DISCOVERY\_USER}};
  \node[state,below= 0.8cm of q6] (q50) {$q_{50}$};
  \node[desc] at (q50.south) {\shortstack{privilege escalation attempt\\\qquad\qquad EMPTY\_NONE}};
  \draw (q0) edge node[edge label] {0.60} (q3);
  \draw (q0) edge node[edge label] {0.05} (q6);
  \draw (q3) edge node[edge label,pos=0.8,xshift=3mm] {0.06} (q18);
  \draw (q6) edge node[edge label,pos=0.7] {0.50} (q50);
  \draw (q18) edge node[edge label,pos=0.7] {0.69} (q26);
  \draw (q26) edge node[edge label,pos=0.85,xshift=2mm] {1.0} (q30);
  \draw (q50) edge node[edge label,pos=0.8] {0.50} (q26);
\end{tikzpicture}
        \caption{Scenario~2: DAG extracted from the learned MC with 57 states.}
        \label{fig:vm2_expl}
    \end{subfigure}
    \hfill
    \begin{subfigure}[t]{0.48\columnwidth}
        \centering
        \begin{tikzpicture}[
  ->,
  >=stealth,
  node distance=0.9cm and 0.8cm,
  every state/.style={draw, rounded corners, thick, fill=gray!5,
                       font=\footnotesize\tt, minimum size=5mm,
                       inner sep=1pt},
  every edge/.style={draw, ->, >=stealth, thick},
  edge label/.style={font=\tiny, fill=white, inner sep=1pt},
  desc/.style={font=\tiny, align=center,
                yshift=-1mm},
]
  \node[state] (q0) at (0.00cm, 0.00cm) {$q_0$};
  \node[desc,xshift=0.3cm,yshift=0.0cm] at (q0.north east) {init};
  \node[state,below=0.4cm of q0] (SCC_5) {$scc_5$};
  \node[desc,xshift=0.4cm,yshift=0.0cm] at (SCC_5.north east) {2 states};
  \node[state,below=0.4cm of SCC_5] (q9) {$q_9$};
  \node[desc,yshift=-1mm] at (q9.south) {\shortstack{enumeration of SUID binaries\\DISCOVERY\_SUID}};

    \node[state,below= 0.8cm of q9] (SCC_2) {$scc_2$};
  \node[desc] at (SCC_2.south) {11 states};
  \node[state,below right = of q9] (q14) {$q_{14}$};
  \node[desc,yshift=-1mm] at (q14.south) {\shortstack{file read (/etc/passwd)\\\qquad INFO\_FILE}};
  
\node[state, below = 0.6cm of SCC_2] (q30) {$q_{30}$};
  \node[desc,yshift=-2mm,xshift=-1mm] at (q30.south) {\shortstack{credential testing\\EXPLOITABLE\_\\CREDENTIAL}};
  \node[state,below right= 0.6cm and 0.8cm of SCC_2] (q33){$q_{33}$};
  \node[desc,yshift=-2mm,xshift=1mm] at (q33.south) {\shortstack{file read \\ (admin\_on\_vac.txt)\\INFO\_FILE}};

  \draw (q33) edge node[edge label] {1.0} (q30);
  \draw (SCC_2) edge node[edge label] {0.89} (q30);
  \draw (SCC_2) edge node[edge label] {0.10} (q33);
  \draw (q14) edge node[edge label] {0.14} (SCC_2);
  \draw (q9) edge node[edge label,pos=0.65] {0.30} (SCC_2);
  \draw (q9) edge node[edge label,pos=0.65] {0.15} (q14);
  \draw (SCC_5) edge node[edge label,pos=0.3] {0.50} (q9);
  \draw (q0) edge node[edge label,pos=0.3] {0.20} (SCC_5);
\end{tikzpicture}
        \caption{Scenario~13: Pruned and collapsed model extracted from an MC with 41 states.}
        \label{fig:vm13_expl}
    \end{subfigure}
    \caption{Illustrative explanation models.}
    \label{fig:strategy_examples}
    \Description{Concise explanation models of successful agentic behavior in Scenarios 2 and 13 of the considered benchmark set.}
\end{figure}

\noindent
\textbf{DAG Extraction} from MCs is a simple yet effective method. Given $ol$: (1) determine all paths from the initial state to $ol$, (2) prune states absent from these paths, and (3) remove loops to create a directed acyclic graph (DAG). Fig.~\ref{fig:vm2_expl} shows a DAG for Scenario 2, which includes all shortest paths to success and compresses the model from $57$ to $7$ states. We can see that the agent always starts with exploration, and that the crucial step is reading the bash history.

\noindent
\textbf{Pruning \& Collapsing.}
DAGs may become large when tasks require extensive exploration. In such cases, we propose to slice the learned model parameterized by $k$, the number of outgoing transitions retained per state. It includes the steps (top-$k$ pruning), which keeps only transitions with the $k$ highest probabilities in each state, (slicing), which prunes all states from which $ol$ is unreachable, and (collapsing), which determines strongly connected components (SCCs) and merges all states in an SCC into a single state.

Fig~\ref{fig:vm13_expl} illustrates this method applied to Scenario 13, reducing the model size from $41$ to $7$ states. The condensed model includes two SCCs, one with 2 states and one with $11$ states, which mainly contain exploration activities. One caveat is that important steps may be collapsed into SCCs. We can observe that reading \texttt{admin\_on\_vac.txt} is important to success ($q_{33}$), but it is also included in $scc_2$.

\section{Conclusion}
\label{sec:concl}
As a step toward our vision of making the latent behavioral structure of AI agents explicit, we propose ATLAS, a model-learning framework that combines LLM-driven abstraction with formal automata learning to recover probabilistic behavior models, capturing the behavioral structure of agentic systems. ATLAS demonstrates that agent trajectories can be transformed into compact symbolic behavioral models. More broadly, we argue that agent trajectories should become first-class modeling inputs rather than passive execution logs.
Our case study on penetration-testing agents demonstrates this need clearly: These agents operate in high-stakes environments, navigate under partial observability, and alternate between exploratory and exploitative behavior that remains difficult to understand from raw execution logs.
 The behavioral models learned in our case study reveal loops, branching points dependent on observed information, and successful testing strategies. The experiments on symbolic knowledge transfer demonstrate that the recovered behavioral models are not merely descriptive. They capture actionable knowledge that can be reused across language models with different capabilities for high-level model-based planning. We further demonstrate how concise explanation models can be derived from the recovered behavioral models through model slicing and simplification.

This work contributes a conceptual framing, an initial pipeline, and a case-study demonstration that opens a research agenda at the intersection of model recovery, runtime analysis, and AI-based systems. By treating learned automata as engineering artifacts, we enable comparative analysis across agent releases, support model-based assurance through safety and security monitoring, and auditing of agentic workflows. We invite the MODELS community to further investigate validation methods, abstraction strategies, and model-management techniques for behavioral models of increasingly autonomous AI agents.
\section{Future Plans} 
\label{sec:future}
We show empirically that ATLAS can recover behavior models from agentic AI trajectories for penetration‑testing applications.

\textbf{Demonstrating Generality.}
The most important next step is to \emph{demonstrate the broader potential of this approach to the research community}. To this end, we plan to recover behavioral models in \emph{various application domains of agentic AI} beyond penetration testing, such as web navigation and software engineering, for example, in agentic software testing. Additionally, we will evaluate whether the  \emph{learned behavioral models are understandable and useful to users of AI agents}. We therefore plan to conduct user studies to assess whether recovered behavioral models support the interpretation of factors influencing successful decision-making, as well as the identification of failure modes.

\textbf{Model-Driven Engineering.}
We plan to demonstrate the potential of the proposed approach by investigating further model-driven activities beyond knowledge transfer and the generation of explanation models. As an immediate next step, we will examine \emph{high-level planning based on identified failure patterns}. Beyond transferring knowledge between language models, we see great potential in deducing symbolic knowledge from learned models with the goal of adapting agentic behavior. Aside from planning, we aim to explore formal verification of recovered behavioral models using probabilistic model checking to validate and debug agentic AI systems, both statically and at runtime -- learned behavioral models could be deployed as runtime monitors to detect deviations from expected agent behavior. Demonstrating these broader uses will help establish \emph{learned automata as a general tool for analyzing, validating, and improving agentic AI systems}.

\textbf{Scalability.}
 Long trajectories, diverse and changing tasks, and heterogeneous environments may pose problems for abstraction discovery and model learning. We must define useful behavioral‑model requirements for agentic systems. An important research question concerns what makes a behavioral model useful. Should models provide coarse, complete coverage, or partial, fine-grained detail?

\textbf{Closing.}
Addressing these questions and challenges will require combining insights from model-driven engineering, formal methods, automata learning, machine learning, and symbolic AI. We believe that recovering behavioral models from agent trajectories will enable new approaches to runtime monitoring, symbolic planning, policy recovery, and systematic model management for autonomous agentic systems.

\begin{acks}
GenAI was used to polish text (ChatGPT), refactor the initial implementation, and add features, like analysis scripts (Claude Code).
% \setlength{\intextsep}{0pt}
% \begin{wrapfigure}[2]{l}{0.5cm}
% \vspace{-2pt}
\includegraphics[width=1cm]{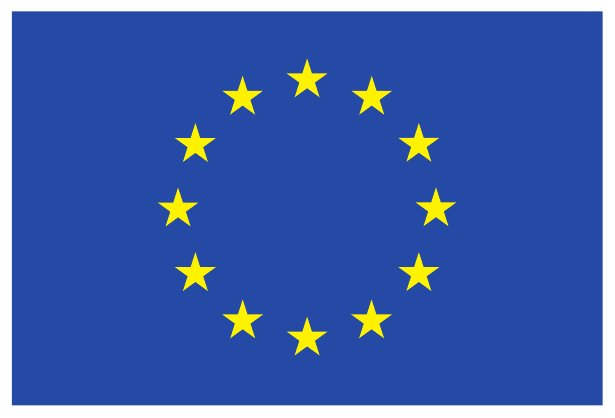}
% \end{wrapfigure}
This work has been supported by
the European Commission in the Horizon Europe research and innovation programme under grant agreements No.
101034440 (MSCA COFUND Doctoral Programme LogiCS@TUWien), No. 101160022 (VASSAL) and No. 101212818 (RobustifAI).

Additionally, it has been partially funded by the Vienna Science and Technology Fund (WWTF) project TAIGER 10.47379/ICT22-023, the FFG-funded project NEST (No. FO999936797), 
and the Austrian Science Fund (FWF) 10.55776/COE12 and 	10.55776/PIN3275223.

\end{acks}

\section*{Ethics and Privacy Statement}
Our case study explores the use of LLMs for penetration testing, which raises concerns about the application of our work by malicious actors. As we build upon existing work in this area, we refer to the original publication describing HackingBuddyGPT~\cite{Happe_2023} for a more thorough discussion of ethical concerns.

\bibliographystyle{ACM-Reference-Format}
\bibliography{references}

\end{document}